\documentclass[aps,showpacs,preprintnumbers,amsmath,amssymb,nofootinbib,preprint]{revtex4}

\usepackage{graphicx}
\usepackage{color}

\def \be  {\begin{equation}}
\def \ee  {\end{equation}}
\def \ee  {\end{equation}}
\def \bea {\begin{eqnarray}}
\def \eea {\end{eqnarray}}

\begin{document}

\preprint{ECTP-2013-02}

\title{Calibrated Fair Measures of Measure:\\
{\small Indices to Quantify an Individual's Scientific Research Output}}

\author{A.~Tawfik}
\email{atawfik@cern.ch}
\affiliation{Egyptian Center for Theoretical Physics (ECTP), MTI University, Cairo, Egypt}

\date{\today}

\begin{abstract}
Are existing ways of measuring scientific quality reflecting disadvantages of not being part of giant collaborations? How could possible discrimination be avoided? We propose indices defined for each discipline (subfield) and which count the plausible contributions added up by collaborators maintaining the spirit of interdependency. Based on the growing debate about defining potential biases and detecting unethical behavior, a standardized method to measure contributions of the astronomical number of coauthors is introduced.

\end{abstract}. 


\maketitle


\section{Introduction}

It is obvious that the publications is the natural outcome of research activity. They are the  announcement about the accomplished activity and achievements to the community of researcher(s). The problem addressed by the present work - the impact of large research collaborations on the publications and citations ascribed to a researcher and the consequences for individual-level evaluations - constitutes an important problem for evaluative scientific production. This is why the topics addressed here have been subjects to discussions in the bibliometrics communities for decades \cite{collabPubs1}. There have been many previous studies of research collaboration  \cite{collabPubs1}. Comparatively, little attention has been given to the concept of {\it research collaboration} or to the adequacy of attempting to measure it through activity (joint projects) and outputs (co-authored publications) \cite{collabPubs1,collabPubs2}. 

In the present work, we suggest to differentiate between collaborations at some levels and classify them according to the coherent preparing and/or release of their publications. We assume that nowadays there are interinstitutional/international collaborations that do not  necessarily involve interindividual and interdependent collaborating researchers. We also show that co-authorship should be a reliable indicator of collaborative teamwork. The co-authorship should reflect essential factors including interdependency. We argue for a more symmetrical and fair approach in comparing the products of collaboration with the disadvantages when considering individual research publications.

The citations are the references added to later publications that subsequently refer to the article of interest. The well-cited publications are recognised as having great impact. But the high citation rates are apparently correlated with other measures of the research excellence. The citations are seen as indicators of the impact of the conducted research. 

The {\it h}-index \cite{hirsh} is  is a scale that attempts to measure both the productivity and impact of the published work of individual or group of scientists, such as a department or university or country, as well as a scholarly journal. It is based on the number of cited papers and their number of citations. A individual or group of scientists or an institution  has index $h$ if $h$ of his/her/its $N_p$ papers have at least $h$ citations each, and the other ($N_p - h$) papers have no more than $h$ citations each \cite{hirsh}. The $h$-index is considered as an alternative to the traditional {\it journal impact factor metrics}. It is demonstrated that $h$ has high predictive value for whether an individual scientist has won prestigious prize. It depends on the {\it academic age} of a researcher. It grows as the citations accumulate.

It is on order now to review other measures. The impact factors is conjectured to suffer from statistical drawbacks. For example, why mean value is favored against the median?
These statistical drawbacks can be counteracted by {counting citation weights fractionally} instead of using whole numbers in the numerators \cite{fract}. 

About fifty five years ago,  Mcconnell argued that {\it ''for anything short of a monographic treatment, the indication of more than three authors is not justifiable''} \cite{sct1,sct2}. We all notice that the number of coauthors kept rising.  It has been recently suggested that: {\it ''in some fields multiple authorship endangers the author credit system''} \cite{sct3}. 

With giant collaborations like Compact Muon Solenoid (CMS) \cite{cms1}, A Toroidal LHC ApparatuS (ATLAS) \cite{atlas1} and A Large Ion Collider Experiment (ALICE) \cite{alice1} at the Large Hadron Collider (LHC) \cite{lhc1}, the number of co-authors of a single paper turns to range from about one to several thousands. For example, $3275$ physicists (among them $1535$ students) and $790$ engineers and technicians are working for CMS. The number of publications continuously raises daily if not more. As per \text{www.inspireshep.net}, there were $2,094$ records from CMS and $2,415$ records from ATLAS registered on  January 14th 2013. On the other hand, when we compare this with a group of $2$ or $3$ thousand scientists working in homogeneous specific fields, we would find that the total sum of their records would exceed that of CMS- or ATLAS-collaboration in the same time interval. Nevertheless, the existing measures of author quality: the mean number of citation per paper, the number of papers published per year \cite{cmp} and the Hirsh's index $h$ \cite{nature} estimate the quality of each author belonging to giant collaborations as extremely higher than that of a non-collaborating author. This might be also valid for some of those few scientists that earn a Nobel prize, whose impact and relevant of scientific work are unquestionable. The first two measures are clear as their names say. Hirsh's index $h$ \cite{hirsh} counts the papers of an individual scientist that have a least $h$ citations each. His remaining papers should have fewer than $h$ citations each.

Besides the given examples on the trends of increasing coauthors' number, we could recall that in $2006$ more than $100$ papers had over $500$ coauthors. There was at least one paper that was signed by $2512$ coauthors \cite{sct4}. Because of the increasing interest in measuring the quantification and estimate the standardization of the scientific impact using various metrics like $h$ index \cite{hirsh,sct7} and because of the growing debate on defining potential biases \cite{sct8,sct9} and detect unethical behavior \cite{sct10,sct11,sct12}, a standardized method to measure contributions of the astronomical number of coauthors is needed \cite{sct1,sct12,sct13,sct14,sct15}.

\section{Unfair discrimination: Disciplines}

Using data from theoretical and experimental high-energy physics available from the inSPIRES database would assure homogeneity of the dataset \cite{lehmann2}. Furthermore, we restrict the comparison between same subfields, either experiment or theory.  For an individual scientist, the number of papers published per year and/or the number citations per paper likely vary from discipline to another. The same is apparently valid for Hirsh's index, $h$.  None of the three measures seems to count for any disciplinary difference. This is one of the largest sources of discrimination. We propose to bring to bear a unitary standard scale charactering each discipline or even subfield, where minimum and maximum marks should be determined. For instance, the minimum and maximum $h$ or number of published papers per year or number of citations per paper should be estimated, regularly. Thus, the determination should be a dynamical process aggregating available dataset and working out the minimum and maximum values according to the utilized measure. The span between minimum and maximum indices could be divided into $100$ equal intervals (bins), for instance. Each bin is supposed to have an equal weight. With this centennial scale, $i$, a main source for discrimination would be to a large extend resolved. A well-known scheme for classification and identification fields and subfields of astronomy and physics  was developed by the American Institute of Physics \cite{aip} and has been used in Physical Review journals \cite{pr} since $1975$ of physics. It turns to a usual process that the authors of scientific papers are required by almost all peer-reviewed international journals to submit about three PACS numbers. Thus, determining measures for each PACS number, $i$, is practically feasible. Such a classification surely offers a quantitative judgement about the contributions of an individual scientist ({\it expertise}) to a specific subfield. The research institutions would find it a helpful tool in their search for the best candidate for a specific position. It would likely happen that an individual scientist makes records in different subfields. The resulting pattern of records in various subfields weights the contributions to each of them.

The abbreviation $i$ is taken from the initial of Imuthes who is considered to be the first engineer, architect and physician in early history. $i$ is nothing but the measures: either the citations per paper or the number of annual papers or the Hirsh's index, estimated for an individual subfield or PACS number. With this regard, other disciplines are invited to develop  classification schemes for their subfields.

\section{Unfair discrimination: Collaboration}

\begin{figure}[htb] 
\includegraphics[width=12.cm,angle=0]{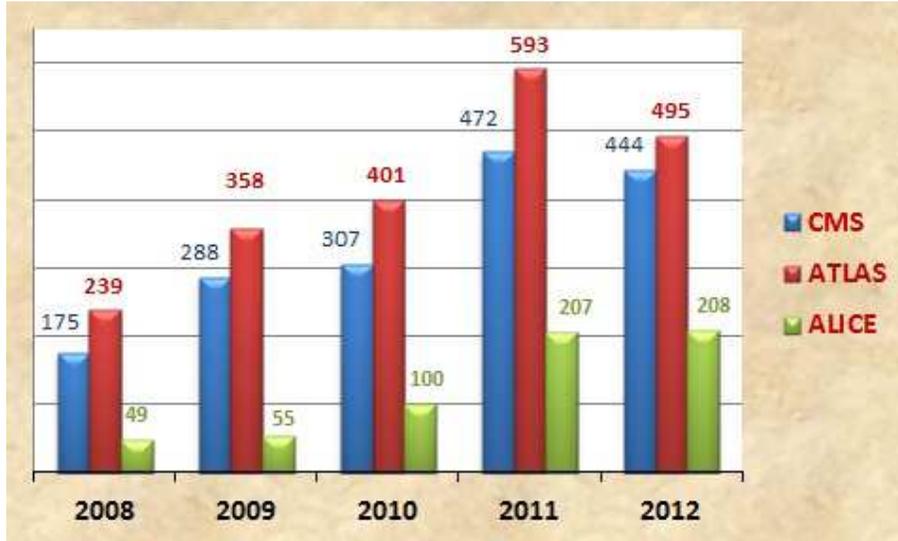}
\caption{{\bf The annual evolution of the number of papers} published by CMS (blue), ATLAS (red) and ALICE (green) collaborations. The slight decrease in 2012 against 2011 is due to not-yet-published papers (records). }
\label{fig:coll}
\end{figure}

To illustrate possible unfair discrimination against the intensive work of thousands individual scientists, we recall inSPIRES data (high-energy physics) for large collaborations in experimental and theoretical physics. Fig. 1 shows the number of annals papers published by the CMS, ATLAS and ALICE collaborations during the last five years. Several hundred papers are brought to publication, annually. Thanks to the generous investments and the intensive work of few thousands of scientists from all over the world on various aspects of physical problems, this huge number of publications are only possible. There is an remarkable increase starting from 2010, coincident with LHC operation. The slight decrease in 2012 compared to 2011 simply originates in the fact that some papers would be in press and probably not yet registered in inSPIRES. To quantify the estimate of an individual's scientific research output, we investigate the number of papers of three scientists, one experimentalist and two theoreticians during the last five years (Fig. 2a). One of the two theoreticians is member of CMS-collaboration since 2008. His record was comparable to that of his colleague  ($72$ compared to $81$). In just five years, his record jumped from $81$ to $353$ making an average increase in the number of papers per year of $54.4$. The record of the other theoretician reflects a reasonable increase (from $72$ to $94$); an annual average increase of $4.4$. Few remarks are now in order. the non-collaborating theoretician is one of the eminent talents. This can be seen from the total number of citations, Fig. 2b. His scientific age is the shortest among the other two. Fig. 2b gives the number of citations of the three scientists in the same time interval. The CMS theoretician gets a remarkable record of citations (from $2116$ to $10140$). Apparently, the benefits of being member of a large collaboration like CMS are also reflected in the Hirsh's index. For the non-collaborating scientist, $h$ increases from $53$ to $61$ in $5$ years, which is about one-third ($h=1.6$) that of the other theoretician joining CMS.

\begin{figure}[htb] 
\includegraphics[width=10.cm,angle=0]{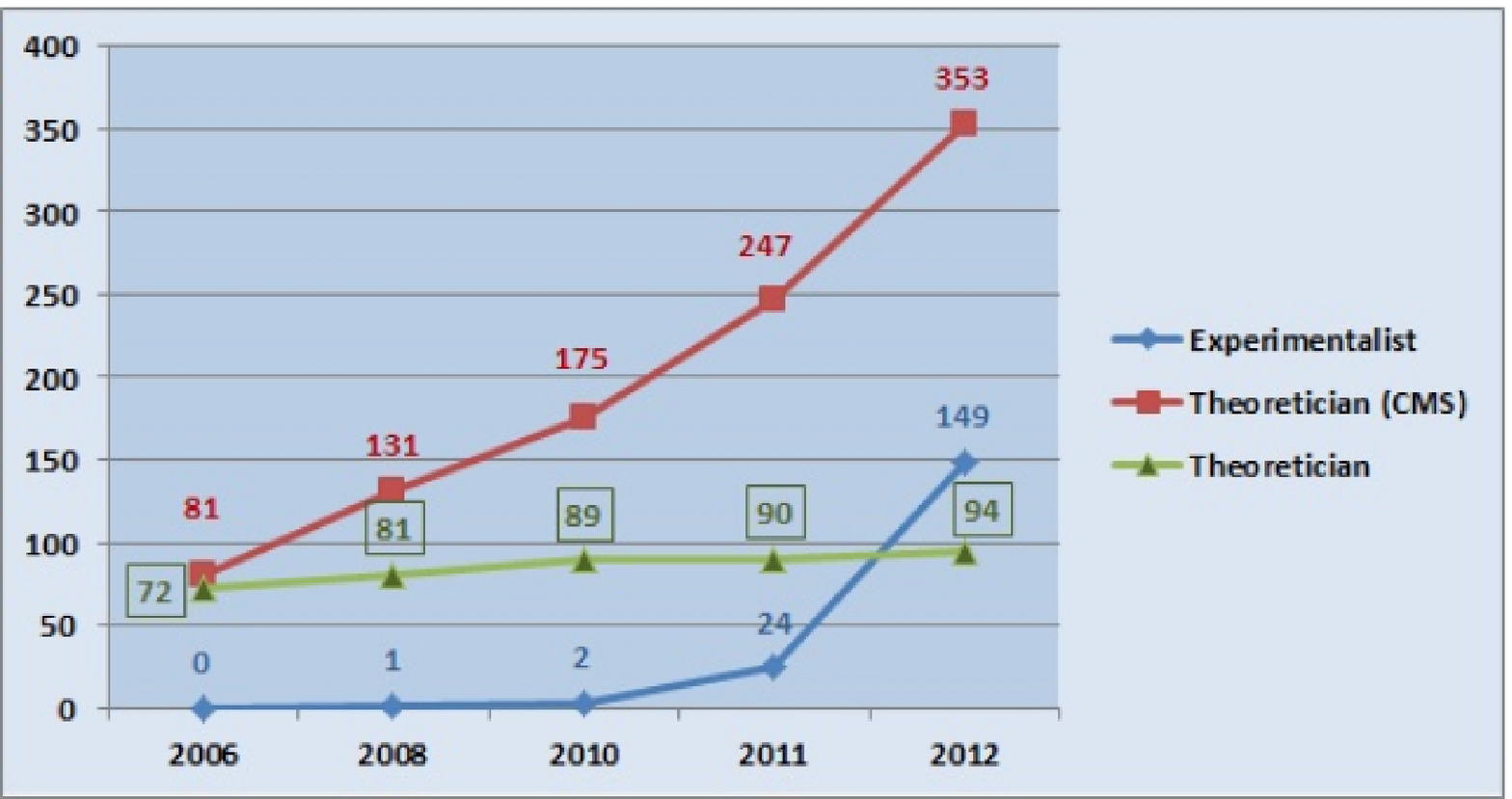}
\includegraphics[width=10.cm,angle=0]{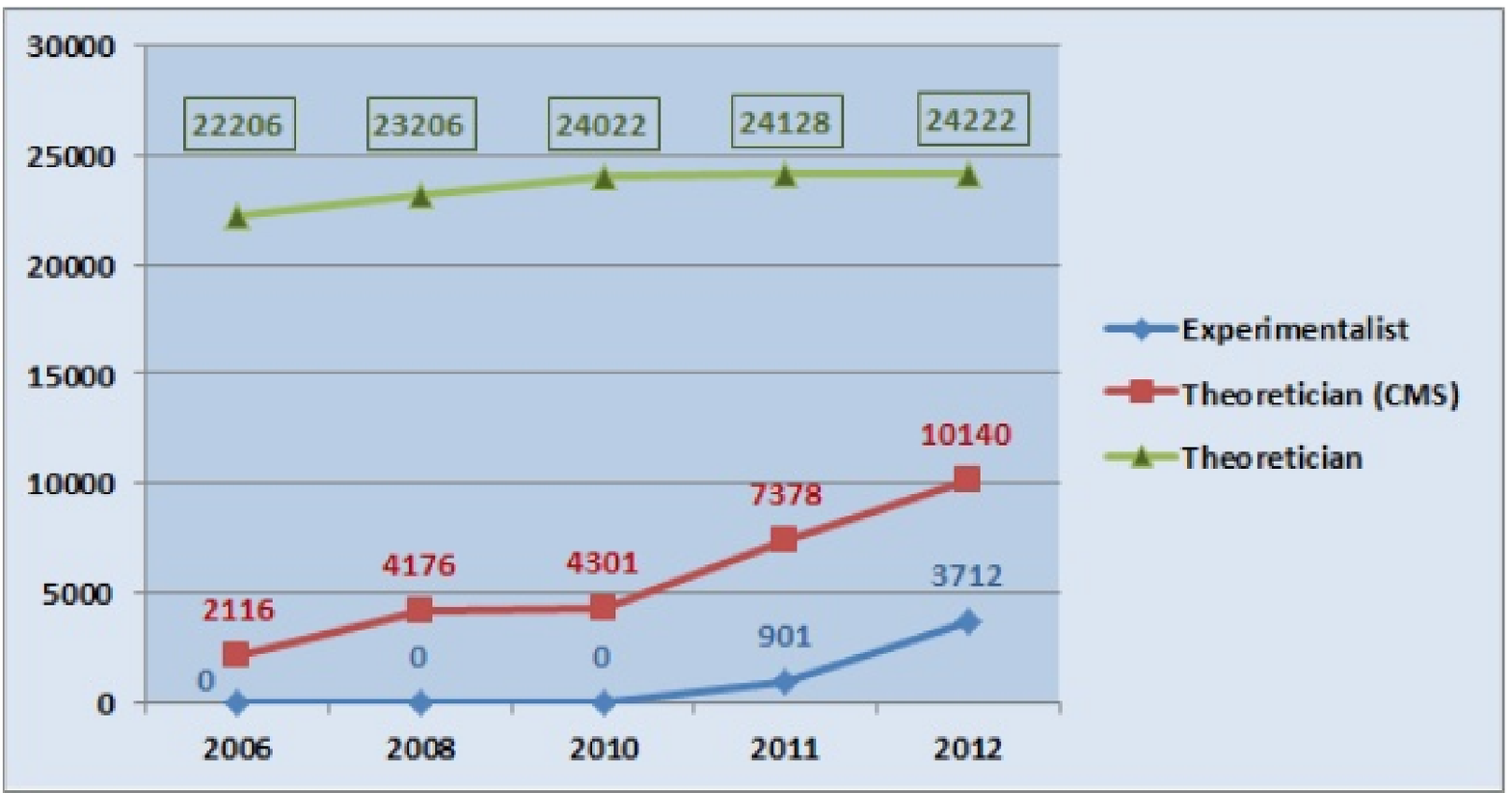}
\includegraphics[width=10.cm,angle=0]{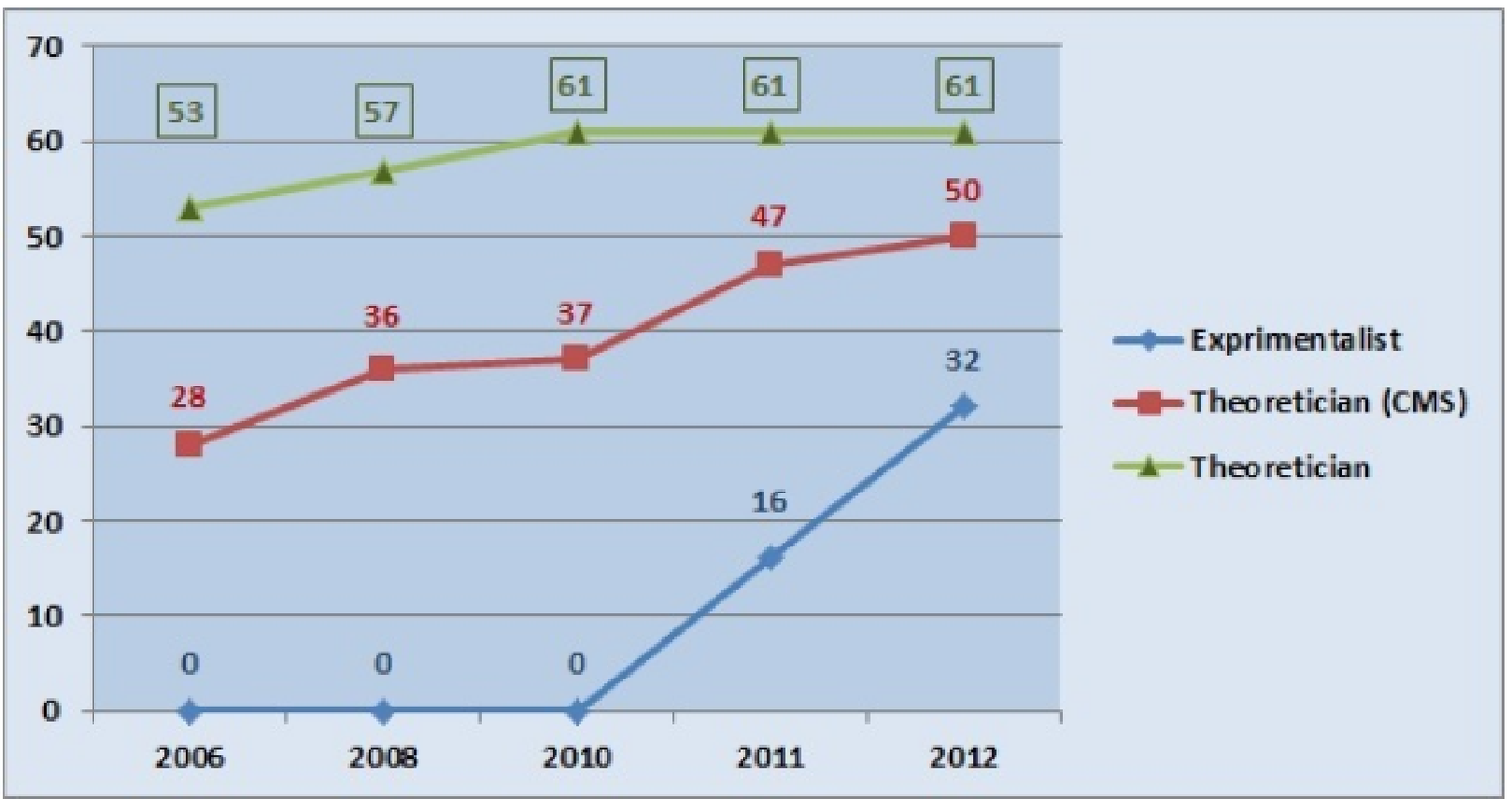}
\caption{{\bf The productivity of three scientists, a-c}, The number of papers per year, total number of citations and Hirsh's index. }
\label{fig:com}
\end{figure}

The experimentalist joined the CMS-collaboration in Autumn 2011. His record of papers jumped from $2$ to $149$ in about $15$ months. The rate of increase is the largest among the other two scientists. His Hirsh's index exploded from $0$ to $32$ in about $1.25$ years.  With this regard, we recall Hirsh's conclusion \cite{hirsh} that in $20$ years 1) a successful, 2) an outstanding and 3) a truly unique scientist can be characterized by $h=20$, $h=40$ and $h=60$, respectively. Accordingly, the Hirsh's parameter $m$ for the three scientists can be calculated in $5$ years (as CMS started couple years ago, this time interval is a quarter of that suggested by Hirsh) as follows. For the experimentalist, $m=6.4$. For the theoretician who is member of CMS, $m=4.4$. For the non-collaborating theoretician, $m=1.6$. It is remarkable that only the third scientist has $m$ parameter that fits with Hirsh's conclusion (between $2$ and $3$). It refers to an outstanding scientist. The $m$ parameters of other two scientists (much greater than $3$) appear {\it out-off-scale}. Their values might not be achieved even by the Nobel Laureates.

In light of the above discussion, the fair measures for the scientific output of an individual scientist has to be calibrated. It is apparent that joining large collaborations brings great advantages against individual scientists or teams of few scientists. An exact estimation for the time span needed to perform a scientific research: formulating the scientific problem, consider all available literature and define own contributions, writing down a paper and finally publishing it, is a complicated process. An average estimation would be deduced from large homogeneous ensemble of annual papers and authors. Homogeneity would refer to certain subfield and/or active authors with a nearly equal scientific age, etc. Averaging the numbers of papers per year per author would help in estimating the time span of producing a paper per author. For completeness, we mention that averaging papers per author is assumed to have {\it rms} variation \cite{nature}. Having such an estimation clarifies that no one would be able to keep an overview at a few hundred papers per year rather than to produce all of them. 

Management of a large collaboration is a sophisticated process. As we are interested in the scientific publication, we roughly describe the utilized algorithm. A central physics committee suggests physical tasks, the analysis is assigned to various small subgroups, solving the physical problems and formulating the text are distributed among other subgroups, then all members are allowed to referee (critics, suggestion, modifications, etc.) the proposal and finally the physics committee takes the responsibility to bring the script to publication. Practically, each subgroup would have a few members. This would simply lead to quenching the entire work to a limited number of members, while others would not or only occasionally participate in, especially when dealing with thousands of members. The spirit of the scientific research is the academic freedom. We inherited this culture and want to deliver it to next generations. It prevents collaboration management from dictating scientific ingenuity to the members. 

The behavior that some members or subgroups out of a huge collaboration come up with most of the work, while others do not, would find its roots even in social sciences.  It is assumed that collaborators should be interdependent and therefore the justification for assigning equal weights to all co-authors would be a subject of further discussion. For instance, we made the experience that authors do not contribute {\it equally} to their papers. The equal contribution can not fairly measured. This assumption is in some disciplines fairly fulfilled and reflected in stable patterns of author sequences (e.g. research conducting most of the experimental work being first author, leader of the group/lab being last author). Fractional counting with equal shares is therefore not an ideal solution. In the present work, we just assume that all co-authors should come up with contribution to their final product, the publication.

Among large groups (people or scientists as well) the interdependency likely falls down (weakens) with increasing the membership. In Sociology, the interdependence describes a relationship in which each member is mutually dependent on the others, which simply differs from a dependence relationship, where some members are dependent, but the rest are not. Therefore, the interdependency is an essential concept in order to deduce the contribution of an individual scientist among a large collaboration. If the collaboration is interdependent, then the contributions of each member can be weighted, equally. Another explanation that no individual scientist whether he/she is member of a large collaboration or not can produce multiple times the averaged number of papers per year per author, would be based on the various topics that are covered by large collaborations.  No individual scientist can be active in all topics so that he/she can produce scientific papers. 

Maintaining the spirit of team work, especially the interdependency in research teams consisting of few {\it homogeneous} scientists, we propose $n$ being an averaged number of maximum {\it real} authors (or maximum number of interdependent team members). Then, any {\it measure} either the citations per paper or the number of annual papers or the Hirsh's index, gets a correction as follows. 
\bea
\text{calibrated measure} &=& measure \left(1- \frac{N-n}{N}\right),
\eea
where $N$ is the collaboration's members. The fact that $N$ should be greater or equal to $n$ means that the above expression is not applicable for single authors and interdependent homogeneous authors. As the interdependency varies from team to team and from collaboration to another, there is no golden number for $n$. One would think to make use of the new standard for literature's list of co-authors. If the list includes more than 10 names, the rest can be replaced by {\it et al.} With collaboration we mean either theoretically or experimentally oriented research or both that jointly conducted by a group of scientists.




\end{document}